Highlights

- Dust particles sizes, optical depth and top altitude maps of a local dust storm
- Evidence for connections between the storm dynamics and the region topography
- Hints on the storm origins from thermal inertia, GCM and surface mafic features

# Properties of a Martian local dust storm in Atlantis Chaos from OMEGA/MEX data

**F. Oliva**, (fabrizio.oliva@iaps.inaf.it), A. Geminale, E. D'Aversa, F. Altieri, G. Bellucci, F.G. Carrozzo, G. Sindoni, D. Grassi, *Istituto di Astrofisica e Planetologia Spaziali, Rome, Italy*.

**Abstract**

In this study we present the analysis of the dust properties of a local storm imaged in the Atlantis Chaos region on Mars by the OMEGA imaging spectrometer on March 2$^{nd}$ 2005. We use the radiative transfer model MITRA to study the dust properties at solar wavelengths between 0.5 μm and 2.5 μm and infer the connection between the local storm dynamics and the topography.

We retrieve maps of effective grain radius ($r_{eff}$), optical depth at 9.3 μm ($\tau_{9.3}$) and top altitude ($ta$) of the dust layer. Our results show that large particles ($r_{eff}$ = 1.6 μm) are gathered in the centre of the storm (lat=33.5° S; lon=183.5 W°), where the optical depth is maximum ($\tau_{9.3} > 7.0$) and the top altitude exceeds 18 km. Outside the storm, we obtain $\tau_{9.3} < 0.2$, in agreement with the estimates derived from global climate models (GCM).

We speculate that a low thermal inertia region at the western border of Atlantis Chaos is a possible source of the dust storm. Moreover, we find evidence that topography plays a role in confining the local storm in Atlantis Chaos. The vertical wind component from the GCM does not provide any hint for the triggering of dust lifting. On the other hand, the combination of the horizontal and vertical wind profiles suggests that the dust, once lifted, is pushed eastward and then downward and gets confined within the north-east ridge of Atlantis Chaos.

From our results, the thickness of the dust layer collapsed on the surface ranges from about 1 μm at the storm boundaries up to more than 100 μm at its centre. We verify that a layer of dust thicker than 1 μm, deposited on the surface, can prevent the detection of mafic absorption features. However, such features are still present in OMEGA data of Atlantis Chaos registered after the storm. Hence, we deduce that, once the storm is over, the dust deposited on an area larger than the one where it has been observed.

*Keywords*: Mars
        Mars, atmosphere
        Dust properties
        Local dust storm
        Imaging spectroscopy

## 1. Introduction

The study of suspended dust on Mars is fundamental to understand the planet's thermal structure and climate (Kahre et al., 2008). Martian aerosols are mainly composed of micron-sized particles, probably produced by soil weathering (Pollack et al., 1979; Korablev et al., 2005), and are composed by nanophase ferric oxide particles (Morris et al, 2006, Poulet et al., 2007). In the visible-near infrared (VIS-NIR) spectral range, these particles contribute to the heating of the lower atmosphere (Korablev et al., 2005). Whereas, in the infrared (IR) domain, they efficiently radiate heat to space through IR emission (Gierasch and Goody, 1972; Pollack et al., 1979; Korablev et al., 2005; Määttänen et al., 2009), strongly influencing the atmospheric circulation, the climate and the weather of Mars. Furthermore, the radiative effects of dust depend on size, composition and on the optical properties of its particles (Pollack et al., 1979; Ockert-Bell et al., 1997; Clancy et al., 2003; Wolff et al., 2003; Vincendon et al., 2007; Määttänen et al., 2009). For these reasons, dust represents a major source of heating and contributes in generating vertical and horizontal winds.

Different dust phenomena happen on Mars, including dust devils (localized), local dust storms (regional) and planet-encircling dust events (planetary-scale) (Briggs et al., 1979; Cantor et al., 2001).

Several mechanisms can trigger the lifting and entrainment of dust particles within dust devils. These are related to both the ambient conditions and the physical characteristics of the dust devil itself (Neakrase et al., 2016), including its orientation with respect to the Sun.

Dust storms usually originate in the Southern Hemisphere. Occasionally, multiple dust lifting sources combine together and grow in planet-encircling storms that envelop the entire planet (Smith, 2004; Cantor (2007); Montabone et al., 2015). For example, Strausberg et al. (2005) and Cantor (2007) suggested that the planet-encircling storm in 2001 started as several distinct local storms grown in the southwest of Hellas Basin, the deepest giant impact crater on Mars. Furthermore, using a Martian atmospheric General Circulation Model (GCM), Ogohara and Satomura (2008) found that the dust of the storms moves northwards driven by the southerly wind.

Doutè (2014) used the data from the Observatoire pour la Minéralogie, l'Eau, les Glaces et l'Activité (OMEGA, Bibring et al., 2004; see Section 2) on board the Mars Express (MEx) ESA spacecraft, to investigate the dust evolution in relation with the seasonal $CO_2$ ice regression in the Southern Hemisphere during the spring of Martian year 27. The author found that there are two mechanisms for lifting and transporting efficiently mineral particles and create dust events or storms: 1) night-time katabatic winds at locations where a favourable combination of frozen terrains and topography exists; 2) daytime mesoscale thermal circulation at the edge of the cap, when the defrosting area is sufficiently narrow.

In the solar spectral range, where solar light reflection prevails the thermal emission, a dust storm usually appears bright and can mask the underlying low-albedo terrains. This behaviour has been verified by analysing OMEGA data (e.g. Vincendon et al., 2015). The same result (Wolff et al., 2009) has been obtained from the analysis of the data from the Compact Reconnaissance Imaging Spectrometer for Mars (CRISM, Murchie et al., 2007) on board the Mars Reconnaissance Orbiter (MRO).

The top altitude of local storms varies significantly in the literature. For example, the thickness and altitude of a bright dust haze in Valles Marineris were retrieved (Inada et al., 2008) using data by OMEGA and from the High Resolution Stereo Camera (HRSC; Neukum and Jaumann, 2004; Jaumann, 2007), both on board MEx. For this haze, formed in mid northern spring, the authors estimated a top altitude of 1-2 km above the bottom of the canyon. On the other hand, Määttänen et al. (2009) studied a different local dust storm event northwards of Pollack crater finding a top altitude of about 10-20 km.

In this work, we study a local dust storm observed on Atlantis Chaos region during Martian Year (MY27) by the OMEGA spectrometer on board MEx. Section 2 describes the instrument and the data set. The method of the analysis and the details on the retrieval model are given in Section 3 and in its subsections. The results of the analysis are presented in Section 4 and discussed in Section 5. Finally, our conclusions are given in Section 6.

## 2. Instrument and observations

OMEGA is the imaging spectrometer on board the MEx spacecraft, orbiting around Mars since late December 2003. The instrument Instantaneous Field Of View (IFOV) is 1.2 mrad and the corresponding spatial resolution ranges from about 0.36 km up to 5 km, depending on the position of the spacecraft on its elliptical orbit (pericenter at about 300 km, apocenter at about 4000 km). OMEGA covers the 0.36-5.10 µm range using the three spectral channels VNIR (0.38-1.08 µm), SWIR (0.92-2.69 µm) and LWIR (2.52-5.08 µm). The spectra are sampled in 352 bands with a spectral resolution of 7 nm, 15 nm and 20 nm for the VNIR, SWIR and LWIR channels respectively. The instrument absolute radiance has a systematic uncertainty of less than 10%, increasing up to 15% at the edges of the spectral channels (Carrozzo et al., 2012; Määttänen et al., 2009).

The data we analyse in this paper are calibrated in radiance and the reflectance factor (RF) is then derived using the geometric information available in the OMEGA geometric files. The local storm has been observed by OMEGA in March 2005 (orbit number 1441_5, Ls=168.619°). In the observation, the storm is cut at west and we cannot see it in its entirety (Figure 1). The part of the storm that we observe is centred at longitude = 176.5° W and latitude = 33.5° S (see Table 1), 13.40 local time. In the solar part of the spectrum at about 1.2 µm the storm appears as a bright cloud and is characterized by more than twice the signal of surrounding regions (Figure 1a). On the other hand, in the thermal range at 5 µm the storm appears as a smooth dark region, suggesting temperatures lower than the surroundings (Figure 1b). Two possible scenarios may explain these colder temperatures: 1) the dust cloud is optically thick both at solar and thermal wavelengths. Hence, only the colder atmospheric layers above the altitude at which it becomes completely opaque contribute to the observed radiance; 2) the cloud is optically thick in the solar range but thin at thermal wavelengths. In this case, the dark region is actually the surface cooling due to the shadowing from the dust. The identification of which scenario is correct would require an extended analysis of the observation at thermal wavelengths and this is beyond the scope of the paper. Nevertheless, both options indicate that the cloud is optically thick at solar wavelengths.

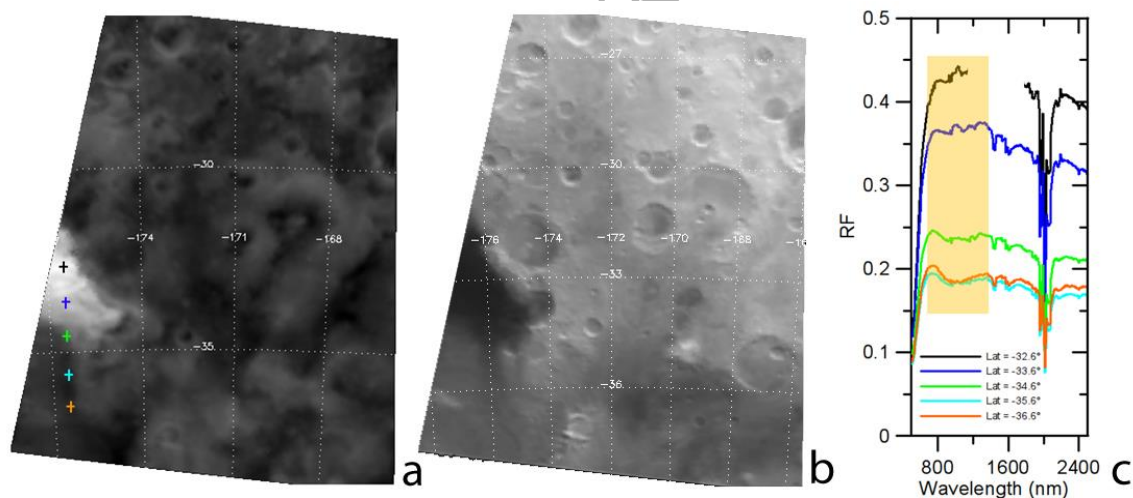

**Figure 1:** *Panel a*: cylindrical map projection of orbit 1441_5 displayed at 1.2 µm. The coloured crosses indicate the locations of the spectra in panel c, selected for the preliminary analysis (see Section 2). *Panel b*: the same data cube displayed at 5 µm. *Panel c*: comparison of the RF at the selected locations, whose latitudes are reported in the legend. The orange rectangle highlights the changing of the mafic absorption band from the surface underneath the cloud. A spectral range affected by instrumental saturation has been removed from the black curve.

Despite the very bright appearance of the cloud in the solar range, no water ice is detectable in this observation. Indeed, water ice diagnostic bands at 1.5 and 2.0 µm are not present and the 3-4 µm spectral range does not show ice features. Moreover, we have computed the Ice Cloud Index (ICI, Langevin et al., 2007), namely the ratio between the RF at 3.40 µm and 3.52 µm, that allows the detection of water ice clouds when its value is lower than 0.8. We have verified that the ICI is higher than 0.8 in the whole observation.

In order to interpret the changes of spectral reflectance between the cloud centre, its boundaries and adjacent off-cloud regions, we have studied a latitudinal scan of 5 pixels of orbit 1441_5 along longitude = 176.5° W and between latitudes 32.6° S and 36.6° S, 1° apart (see the crosses in Figure 1a). By comparing the 5 spectra in the wavelength range 0.5-2.5 μm (Figure 1c), it is evident how the reflectance decreases by going from the centre to the boundaries of the storm, and then to the off-storm regions. This is a hint that the dust optical depth is decreasing along the scan (see Section 3.1). In particular, it is interesting to note that the spectral shape changes between around 0.7 and 1.4 μm (see the orange rectangle in Figure 1c). This happens since the surface mafic absorption signature becomes more and more important along this track, further suggesting a marked decrease of the dust optical depth. As a consequence, an independent knowledge of the spectral albedo of the surface underneath the cloud is required in order to analyse the storm properties with a radiative transfer (RT) model (Section 3).

In order to define the properties of the surface underneath the dust cloud, we search the whole OMEGA dataset for observations of the same region with lower dust content. Such observations allow the retrieval of the actual surface albedo.

Table 1 summarizes the most significant parameters of the selected candidates to perform this retrieval. The details of orbit 1441_5, where the storm has been observed, are also given. With the exception of orbit 1474_5, the ICI is always above 0.8 in the cloud region and, therefore, we can rule out the possible presence of water ice clouds. Moreover, among all, orbits 3262_5 and 5654_5 show a lower atmospheric dust content (Montabone et al., 2015). Hence, they are the most suitable to derive the surface properties (see Section 3.1).

| Orbit name | Latitude S [°] | Longitude W [°] | Ls [°] | $\tau_{9.3}$ | ICI | rs [km/pixel] |
|---|---|---|---|---|---|---|
| *ORB1441_5* | *32.6* | *176.5* | *168.6* | *0.18* | *0.92* | *4.3* |
| ORB1258_3 | 32.6 | 176.5 | 141.9 | 0.14 | 0.86 | 1.9 |
| ORB1474_5 | 32.5 | 176.4 | 173.7 | 0.23 | 0.67 | 4.8 |
| **ORB3262_5** | **32.6** | **176.5** | **83.8** | **0.08** | **0.87** | **1.7** |
| ORB5161_3 | 32.6 | 176.5 | 14.5 | 0.11 | 0.93 | 0.9 |
| **ORB5456_5** | **32.6** | **176.5** | **52.9** | **0.04** | **0.87** | **3.3** |

**Table 1: list of all the orbits selected to check the best conditions for the retrieval of the surface properties. Orbit 1441_5, in which the storm has been observed, is displayed in italic. All values refer to the centre of the storm as it is observed in orbit 1441_5 (see Section 2). The optical depth values at 9.3 μm ($\tau_{9.3}$) are derived from Montabone et al. (2015). *Ls* is Mars Solar Longitude; *ICI* is the Ice Cloud Index (Langevin et al., 2007); *rs* is the spatial resolution in km/pixel. Orbits given in bold represent the best candidates to perform the retrieval of the surface properties (see Section 2).**

## 3. Method and model

Our analysis is focused on the retrieval of the dust cloud particles effective radius ($r_{eff}$), top altitude (*ta*) and optical depth at 9.3 μm ($\tau_{9.3}$) in the region of orbit 1441_5 extending between longitudes 176.5° W and 177.5° W and latitudes 32.0° S and 35.7° S. We selected this region because is a part of the intersection between orbit 1441_5 and orbit 3262_5, where the spectral surface albedo can be obtained (see Section 3.1). This region contains also some off-storm pixels that can be diagnostic of the behaviour of the retrieval code outside the cloud. In order to speed up the computations, we have performed a 3x3 pixels binning in the selected region. As a result, the analysis has been performed on a 200 pixels area in which the spatial resolution is about 12 km/pixel. The wavelength at 9.3 μm has been chosen to compare our results with those obtained from global climate models (e.g. Montabone et al., 2015).

We only consider wavelengths between 0.5 μm and 2.5 μm for a total of 189 OMEGA bands. We do not consider shorter wavelengths due to increasing uncertainties in the calibration. On the other hand, longer ones have not been taken into account since in this study we concentrate on the scattering properties of dust, whose information content is enhanced at solar wavelengths. This way, the contribution due to the thermal emission in the RT model can be neglected.

For the analysis we adopt the RT model described in Oliva et al. (2016), Adriani et al. (2015) and Sindoni et al. (2013), already used to study the properties of Saturn's and Titan's clouds and hazes. The forward model is based on the full multiple scattering DISORT solver (Stamnes et al., 1988; Stamnes, 2000) and relies on the assumption of 1D atmosphere in local thermal equilibrium (LTE). Mie theory is adopted to compute the single scattering properties of the aerosols.

The inversion algorithm is based on the Bayesian approach and adopts the Gauss-Newton method to iteratively minimize the cost function (Rodgers, 2000), whose convergence is defined through the reduced $\chi^2$ analysis.

The errors associated to the *a priori* parameters, to the forward model and to the observations are discussed in Section 3.4.

An extended review of the sensitivity tests behind the forward model and the inversion algorithm is given in two appendices at the end of the paper by Oliva et al. (2016).

### *3.1. Surface albedo spectra retrieval*

In Section 2 we have described how orbits 5456_5 and 3262_5 are suitable to retrieve the surface albedo spectra of the storm region, since no water ice clouds are present and the dust opacity is low. Both orbits cover only partially the region interested by the storm, but 3262_5 allows a larger coverage and has thus been used to perform the retrieval of the surface properties. Unfortunately, one half of the 3262_5 observation is affected by strong striping effects and we discard that part.

Orbits 1441_5 and 3262_5 are compared at 1.2 μm in Figure 2a and as RGB maps in Figure 2b. The contour lines in Figure 2a represent the isolines of RF at 1.2 μm, decreasing from the black to the blue line. As abovementioned, this is a hint that the optical depth of the cloud is decreasing from the centre to the boundaries of the storm (see Section 2). Figure 2b shows how suspended dust can be more easily distinguished by using RGB maps, as it appears with a whitish colour with the given RGB combination.

By comparing the maps of the two orbits, it is evident that the sharpness of the surface features in 1441_5 is reduced more than expected with respect to orbit 3262_5, even given the different spatial resolution of the two observations (see

Table 1). This suggests that the dust scattering is enhanced in orbit 1441_5, also outside of the storm region.

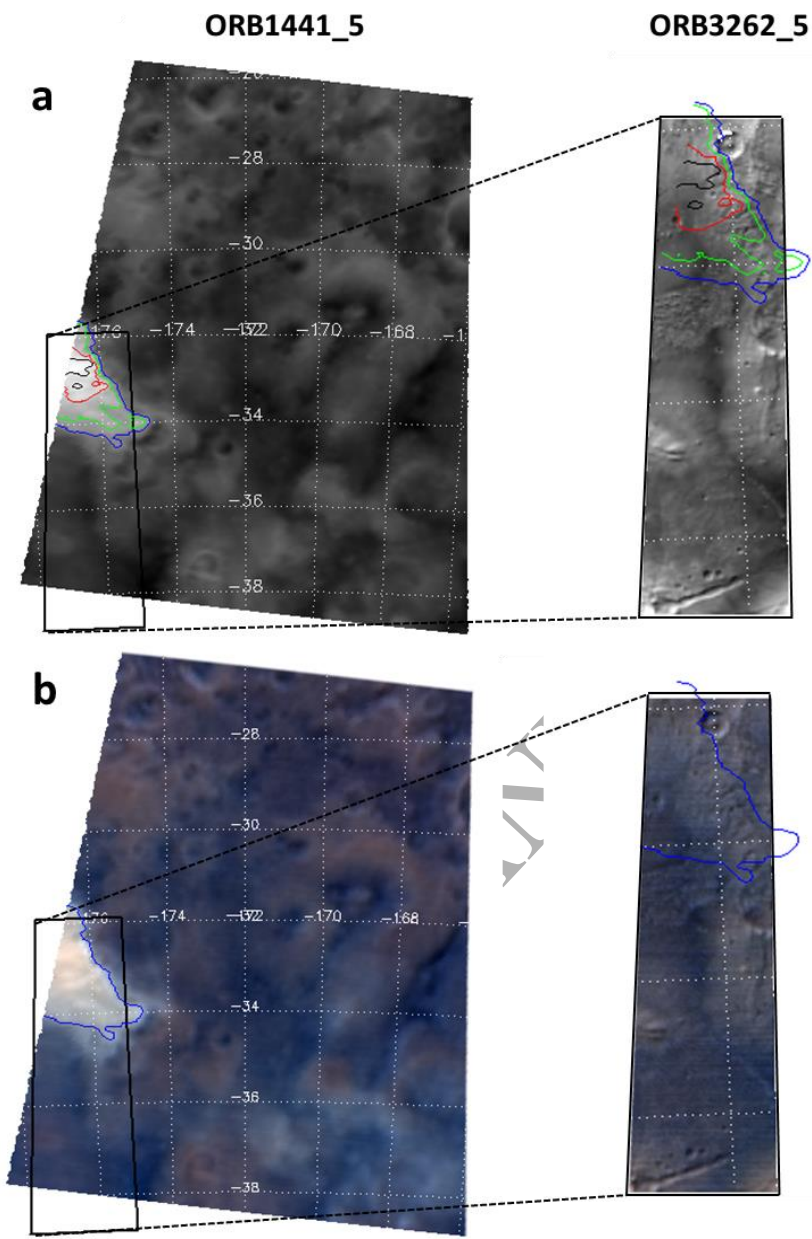

**Figure 2:** *a)* comparison of the RF maps of orbits 1441_5 (left) and 3262_5 (right) displayed at 1.2 μm. One half of the 3262_5 observation has been discarded due to strong striping effects. The thick black rectangle superimposed on orbit 1441_5 map indicates the footprint of orbit 3262_5. The coloured isolines give the 1.2 μm RF of the dust cloud at values of 0.50, 0.40, 0.35 and 0.30 (from black to blue). *b)* RGB maps for the two observations, where R=0.73 μm, G=0.58 μm and B=0.51 μm.

In order to remove the atmospheric contribution and obtain the spectral albedo of the surface, the spectra from orbit 3262_5 have been processed using a statistical method described by Bandfield et al. (2000). This procedure, hereafter referred as Surface Atmosphere Separation (SAS), is based on the principal component analysis (PCA) and Target Transformation (TT) technique (Hopke, 1989) and has already been applied to OMEGA data by Geminale et al. (2015). The PCA allows finding a set of eigenvectors whose linear combinations are able to reconstruct all the original measured spectra. The TT provides the means for physically interpreting the eigenvectors, by associating them to a set of trial vectors representing the possible physical contributions to the measured data. Trial vectors for atmospheric contributions, representing the transmittance spectra of $CO_2$, CO and $H_2O$ gases, have been computed using a simple line-by-line approach. The trial vector describing the surface component has been obtained with the Mons Olympus (MO) method (McGuire et al., 2009). An example of the improvement provided by the SAS method to the surface albedo retrieval is shown in Figure 3a, where a sample SAS-retrieved surface reflectance (thick line) is compared with the corresponding MO-retrieved one (light line). SAS method allows improving the MO method by better describing the gases transmittance. Figure 3b shows the SAS-retrieved gases transmittance (thick line) compared to the MO-retrieved transmittance (light line). The PCA algorithm allows to retrieve the real shape of the gases transmittance, revealing the presence of bad spectral pixels that can produce spikes when MO attempts to obtain the RF of the surface. Four examples of surface albedo spectra obtained with the SAS method are shown in Figure 4.

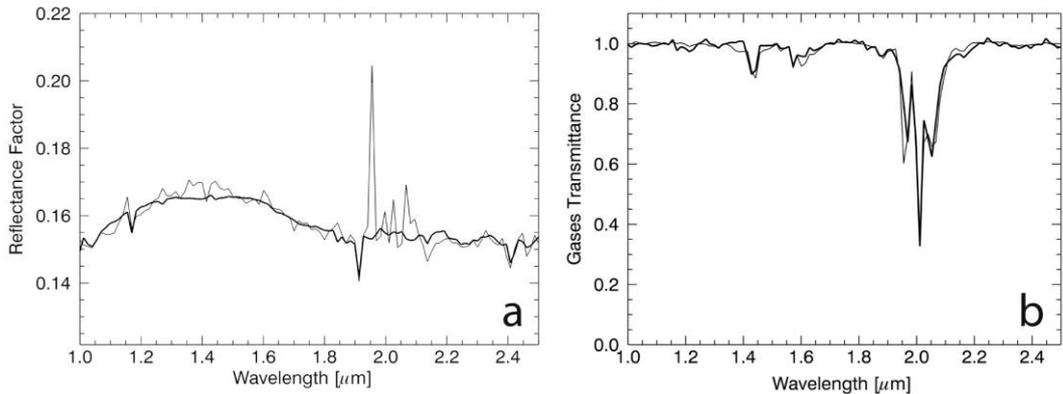

**Figure 3:** *Panel a*: retrieved surface reflectance (thick line) by means of the SAS method (for latitude = 33.6° S and longitude = 176.5° W in orbit 3262_5) compared to the surface reflectance obtained with MO method (light line). *Panel b*: retrieved gases transmittance with SAS method (thick line) compared to the transmittance used by MO method (light line).

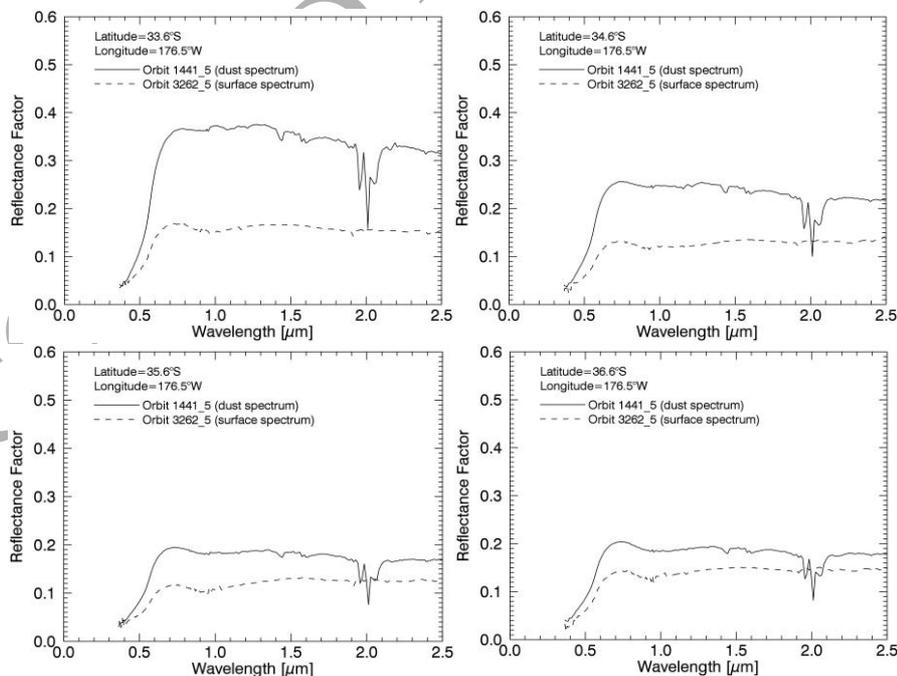

**Figure 4:** comparison between the dust cloud spectra and the underneath surface spectra of the pixels between latitudes 33.6° S and 36.6° S shown in Figure 1. In all panels the thick lines are relative to the dust cloud of orbit 1441_5, while the dashed lines are the surface spectra obtained after applying the SAS method to orbit 3262_5.

## *3.2. The model atmosphere*

We use the Mars Climate Database (MCD, Millour et al., 2015, Lewis et al., 1999; Forget et al., 1999) to compute the dependence of temperature and pressure from the altitude in the atmosphere (T-P profiles) and the mixing ratios of the gases considered in the model. MCD takes into account different dust and solar extreme ultraviolet (EUV) scenarios. We have tested three scenarios to describe the data: scenario 27, related to Martian year 27 that is specific of orbit 1441_5; scenario 7, corresponding to "dusty atmosphere" conditions; scenario 5, corresponding to planetary-scale dust storm conditions. These scenarios also give different mixing ratios for $CO_2$, CO and $H_2O$. Hence, they provide the expected range of variability for the adopted atmospheric gases. Forward models produced with these three cases and with fixed dust optical depth differ of less than 2% in the considered spectral range. Since there is no significant difference in these simulations, we assume MCD scenario 7 for the T-P profiles of all the pixels of the region selected to perform the analysis (see Section 3). We chose scenario 7 among the three because, while scenario 27 is representative of the specific Martian year of orbit 1441_5, it does not take into account the presence of the local storm. On the other hand, scenario 5 is specific for planetary-scale dust storm conditions and is not representative of the local dust storm we study here.

Throughout the region of the storm, the T-P profiles are spatially quite uniform among each other (see Figure 5), except for the lowest atmospheric layers where the changing surface altimetry has a not negligible effect on the simulations. To take into account these differences, we have divided all the surface altimetry values of the selected region in 10 classes, linearly spaced between -0.4 km and 1.8 km. For each class, a representative T-P profile has been selected from the MCD.

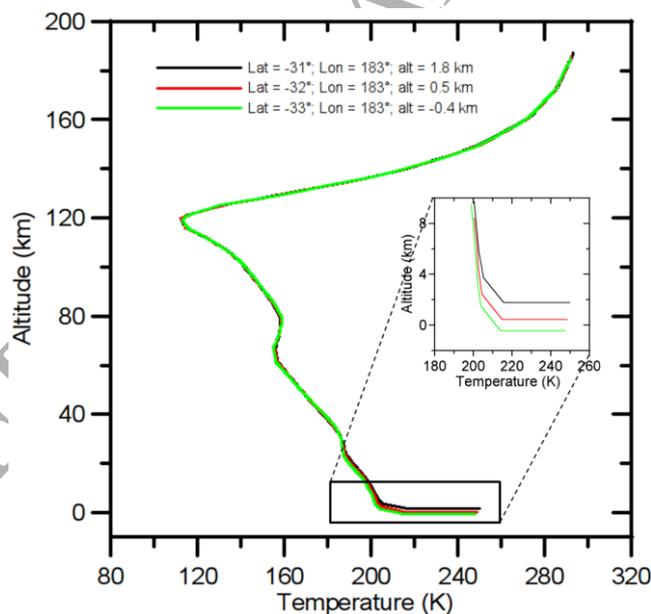

**Figure 5: the main panel shows three T-P profiles (scenario 7, MCD) related to different altimetry values. The window in the middle of the plot shows a zoom of the profiles between -2 and 10 km.**

As in the retrieval of surface albedo spectra (see Section 3.1), we consider $CO_2$, CO and $H_2O$ as gaseous components, well mixed in the atmosphere. The related mixing ratios have been taken from MCD and have been assumed as fixed during the retrieval. Gaseous absorption cross sections have been computed using the HITRAN 2012 database (Rothman et al., 2013) and exploiting the routines of the ARS RT code (Ignatiev, 2005). We have verified that the contribution due to Rayleigh scattering is negligible in the studied range and, for this reason, we do not consider its effects in the model.

### 3.3. Dust geometrical and microphysical properties

We adopt the optical constants of Wolff et al. (2009) to account for the dust composition. Indeed, we verified that they can reliably reproduce the local storm spectra, even if they were derived from data relative to the planet-encircling dust event of 2007.

We assume that the dust cloud is made of a single deck with adjustable top altitude ($ta$) and extending down to the surface. We assume a lognormal functional form to describe the particles size distribution, with a fixed effective variance ($v_{eff}$) of 0.5 (value taken from MCD). Moreover, we assume that the values of $r_{eff}$, $ta$ and $\tau_{9.3}$ in the *a priori* state vector are 1.76 μm (from MCD), 5 km and 1 respectively. The $ta$ and $\tau_{9.3}$ values have been chosen arbitrarily to represent a moderately opaque and extended dust layer.

For each pixel of the studied region (see Section 3), we derive the corresponding surface albedo spectrum, as described in Section 3.1, and we use it as input in the retrieval model.

Figure 6 shows two examples of synthetic spectra best-fitting two observations selected on and off the cloud. It is evident that this model, with a single dust layer extending down to the surface, is able to describe with sufficient accuracy all the spectral signatures related to both the cloud and the surface.

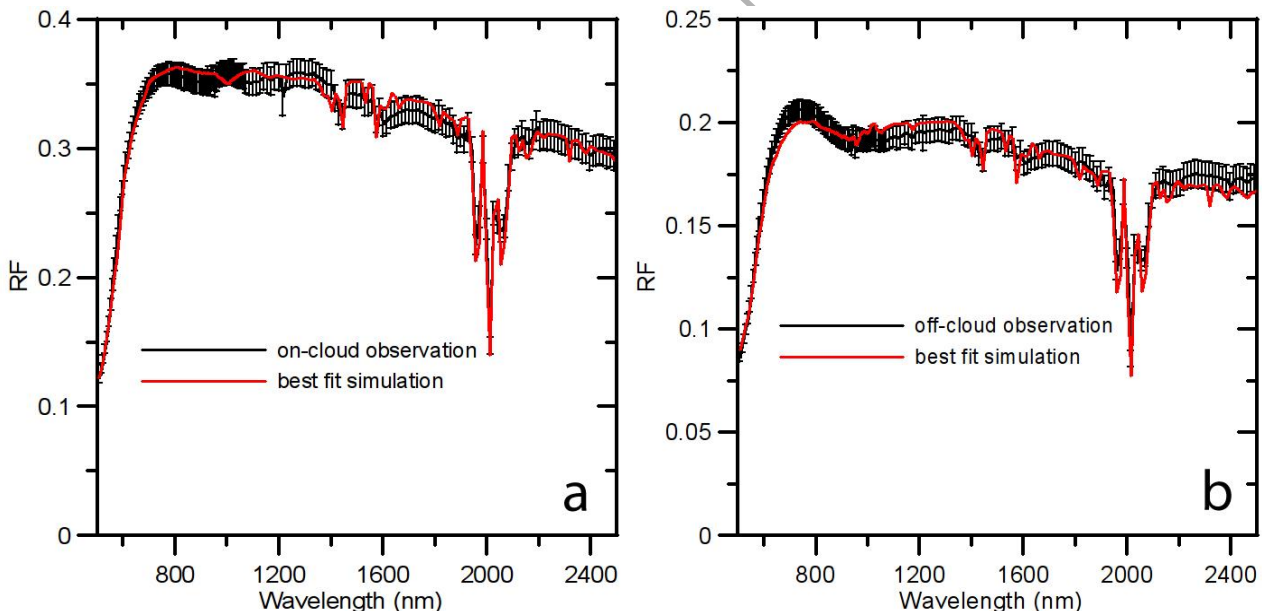

**Figure 6: examples of best fit simulations related to two pixels selected on the dust cloud (a) and off the dust cloud (b).**

The results we have obtained for $r_{eff}$, $ta$ and $\tau_{9.3}$, are presented in Section 4 (see Figure 7).

### 3.4. Errors estimation

In this section we deal with the estimation of the uncertainties related to the observations, the forward model, the *a priori* and the retrieved parameters.

#### 3.4.1. Spectral inversion input errors

The Gauss-Newton iterative method used for spectral inversions requires the definition of covariance matrices of the *a priori* parameters ($S_a$) and of the observation+forward model ($S_\varepsilon$) errors (Rodgers, 2000). Regarding $S_a$, since we used the same *a priori* for both on- and off-cloud spectra, we adopt constraints broad enough to ensure convergence for the whole dataset. Indeed, the wide spectral variations in the observed spectra are indicative of very different radiative regimes and the related parameters may differ significantly from the *a priori* values.

Regarding the measurements noise, on average, OMEGA data have SNR values greater than 100 (Bibring et al, 2004). The major contribution to the uncertainty of the forward model comes from the surface albedo retrieval, estimated to be less than 2% on average in the range 0.5-2.5 μm (Geminale et al., 2015). By taking into account all the other relevant sources of error, i.e. the uncertainty related to the MCD T-P profiles (see Section 3.2), we have estimated that the sum of the observations and of the forward model errors is on average 5% throughout the whole considered spectral range.

### 3.4.2. Retrieved parameters errors

Physically meaningful errors for the retrieved parameters have been computed by means of a statistical method (bootstrapping). We applied this method to the 5 selected spectra (Figure 1c) described in Section 2 and assumed to be representative of the different radiative regimes of the cloud and of the regions outside the cloud itself. The measurement errors are used to create a randomized sample of artificial spectra following the same statistics as the true observations. Then, an error value is derived for each parameter and each observed spectrum by studying the distribution of the parameters retrieved from each sample. We used the sum of the observation and of the forward model uncertainties as error bars in the retrievals, in order to take into account the systematic errors of the inputs of our model.

One remark must be given about the errors related to *ta* for the zones away from the cloud centre. During the retrievals performed on the randomized samples relative to these regions, *ta* always converged to the surface (lowest level of the adopted T-P profile) yielding a 0 standard deviation. For this reason, to compute the *ta* error we take into account the difference in height between the surface and the next adjacent level in the T-P profile. Assuming that this value is equal to 3σ, the *ta* error for the off cloud zones is equal to σ. See Section 4.1 for the interpretation of the results in the off cloud zones.

In Table 2 we give the errors for the 5 regions defined by the coloured contour lines shown in Figure 2a: the *black zone* is comprised within the black line; the *red zone* is centred on the red line and goes from the black to the green line; the *green/blue zone* is the region across the green and the blue lines; *off cloud 1* and *off cloud 2* zones are relative to two off-cloud regions identified by the cyan and orange crosses in Figure 1a.

The low $r_{eff}$ and $\tau_{9.3}$ errors away from the centre of the storm derive from the fact that we are retrieving only 3 parameters from observations containing about 190 spectral points. For this reason, these values have to be considered as averages over the 190 observations contained in each spectrum.

| Parameter | black zone | red zone | green/blue zone | off cloud 1 zone | off cloud 2 zone |
|---|---|---|---|---|---|
| $r_{eff}$ | 7 % | 2 % | < 1% | < 1% | < 1% |
| $ta$ | 2.8 km | 1.3 km | 0.3 km | 0.2 km | 0.2 km |
| $\tau_{9.3}$ | 30 % | 7 % | 3 % | 1 % | 2 % |

**Table 2: errors of the retrieved parameters for each representative region of the storm. Each zone is indicative of areas defined by the coloured contour lines shown in Figure 2a (see Section 3.4.2 for details).**

We also estimated the effects of systematic calibration uncertainties on the variability of the retrieved parameters. OMEGA's absolute photometric calibration is reliable down to 10% (see

Section 2). To take into account this effect we have performed the retrieval on two case spectra relative to high and low dust loading conditions. For each spectrum, the retrieval has been performed 3 times: one with the nominal signal and other two by increasing and decreasing respectively the observed spectrum by 5% (to ensure a 10% total variability). Then, we computed the standard deviation of the results for each parameter to estimate how much they vary due to this effect. The results are shown in Table 3.

| Case | $r_{eff}$ variability | ta variability | $\tau_{9.3}$ variability |
|---|---|---|---|
| High $\tau_{9.3}$ | 6 % | 6 km | 27 % |
| Low $\tau_{9.3}$ | 1 % | 0 km | 9 % |

**Table 3: variability of the retrieved parameters due to systematic uncertainties in the radiometric calibration, estimated for high and low dust loading conditions.**

The 0 km variability for the low optical depth case comes from the fact that the retrieval always converged to the same level of the T-P profile.

The values in Table 3 should be considered aside from the errors we computed in Section 3.4.2. Indeed, those errors are relative to the fluctuations of the observations and to the uncertainties of the forward model. On the other hand, the values in Table 3 indicate how our results would change if the observations had a different calibration and, hence, should not be considered as errors on the parameters.

## 4. Results

In this section we show and discuss the quantitative results that we have obtained with our retrieval. We compare the spatial maps of the dust $r_{eff}$, ta and $\tau_{9.3}$ with the topography of Atlantis Chaos to investigate possible connections with the storm dynamics (Section 4.1). As side products of our analysis, we also discuss the dust column density and its precipitable thickness (i.e. the thickness of the dust layer collapsed on the surface), both computed from $r_{eff}$ and $\tau_{9.3}$.

### 4.1. Retrieved parameters maps

Figure 7 shows the projected maps of orbit 1441_5 and of all the retrieved and computed parameters (see Section 4), superimposed on a MOLA Shaded Relief map (Neumann et al., 2001) of the Atlantis Chaos region.

The RF map at 0.9 μm of orbit 1441_5 is displayed in panel *a*. The storm appears to be confined within the north-east ridge of Atlantis Chaos, climbing it over the 34° S parallel. The dust cloud splits in two arms eastward of the 176° W meridian: the southern arm resides at the bottom of the crater while the northern one elongates along the north-east ridge. The map shows how the storm is probably larger than the part we see here and probably extends on the west side of Atlantis Chaos. The cloud optical depth ($\tau_{9.3}$) is shown in panel *b*. As already suggested in Section 2, the dust opacity is maximum at the centre of the storm ($\tau_{9.3} > 7.0$) and decreases proceeding towards the cloud boundaries ($\tau_{9.3} = 1.0$). The optical depth map reproduces the two arms already identified in panel *a*. These are characterized by a similar optical depth ($\tau_{9.3} \sim 2.0$). Outside the dust cloud, the opacity suddenly drops to values less than 0.2, in agreement with the estimates of Montabone et al. (2015) (see Table 1). This also happens towards the Magelhaens crater (blue dashed line) beyond the north-east ridge of Atlantis Chaos, suggesting again that the dust cloud is somehow confined by the topography.

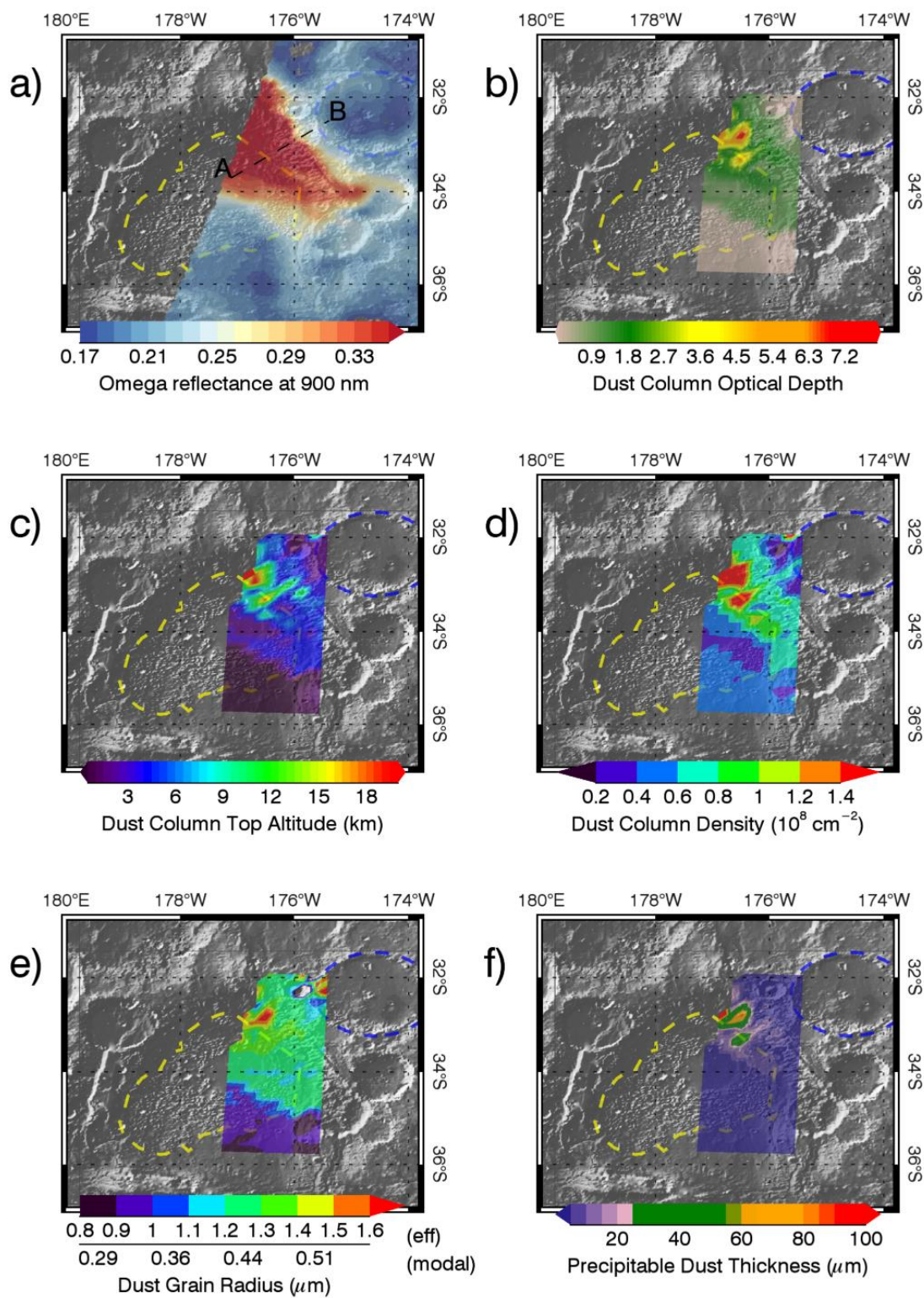

**Figure 7: projected maps of orbit 1441_5 (panel *a*) and of the retrieved and computed parameters (panels *b* to *f*) superimposed in transparence on MOLA Shaded Relief maps (Neumann et al., 2001). The yellow dashed line indicates the border of the Atlantis Chaos region; the blue dashed line indicates the border of the Magelhaens crater.**

The cloud top altitude (*ta*) is shown in panel *c*. The maximum altitude of the dust column exceeds 18 km at the centre of the storm. As in Figure 7*a* and Figure 7*b*, also the *ta* parameter reproduces the two arms of the cloud. The northern one has *ta* higher than 12 km while the southern one does not exceed 10 km. At the storm boundaries *ta* decreases down to a maximum of 4 km while outside the cloud it is less than 1.5 km. Here, the results indicate that the dust layer is compressed on the surface and is characterized by a low optical depth. However, it is worth stressing that in this region the cloud could be either compressed on the surface (as in our results) or vertically more extended but with the same low columnar optical depth. Indeed, for such low values of $\tau_{9.3}$, the measurements do not contain enough information about the actual vertical profile of dust density and our model is only sensitive to its integrated opacity. Due to the lack of information for the cloud top altitude, in this region our results converge to the compressed scenario since *ta*, in the *a-priori* state vector, is closer to the surface than to the top of the atmosphere. For these reasons, we cannot constrain *ta* in the off-cloud zones. The *ta* results and errors (see Section 3.4.2) we obtain in these regions should be considered under the above considerations.

The dust column density, derived from the retrieved optical depths, is shown in Figure 7*d*. As in the case of the optical depth and top altitude maps (Figure 7*b* and Figure 7*c*) this parameter has maximum values at the centre of the cloud where we find more than $1.5 \times 10^8$ particles/cm$^2$. The two arms of the cloud are reproduced once again and the southern arm is denser ($10^8$ particles/cm$^2$) than the northern one ($8 \times 10^7$ particles/cm$^2$). This behaviour will be discussed in Section 5.

The particles dimension map is shown in panel *e*. This map shows larger particles at the storm centre ($r_{eff}$ = 1.6 μm). Then, the dimensions decrease by proceeding radially from the centre to the boundaries of the cloud ($r_{eff}$ = 1.2 μm) and no arms can be detected.

Finally, we also computed the map of the precipitable dust thickness (Figure 7*f*). This is the thickness of a hypothetical dust layer if all the cloud spherical particles would collapse on the surface with a close-packing arrangement. This quantity, derived from the retrieved grains size distributions and from the computed column densities, yields values below 1 μm outside the cloud and above 20 μm on the cloud, reaching more than 100 μm at the centre.

## 5. Discussion

In Section 4.1 we have presented the quantitative results obtained from the retrieval. We have seen how the dust cloud appears to be confined by the topography of the Atlantis Chaos region. Indeed, by crossing the north-east ridge towards lower latitudes and longitudes, all the parameters drop to lower values (Figure 7). The cloud confinement is better appreciated in Figure 8, showing a vertical section of the retrieved results along the transverse track indicated by the black dashed line in Figure 7a. It is evident how the storm extends to high altitudes at the centre of the crater, where the particles dimensions are large and the dust content is enhanced. Then, departing from the centre, the top altitude and the density decrease almost symmetrically. In particular, the cloud practically disappears when it crosses the north-east ridge of Atlantis Chaos.

While a mesoscale dynamical model could help in understanding where the storm originated and where it is heading, such an analysis is beyond the scope of this paper. However, we discuss the storm origins and outcome with considerations on thermal inertia, winds and on the detectability of the surface mafic absorption feature around 1 μm.

The retrieved dust distribution is compatible with a confinement of the storm at its eastward boundary due to the topography. This hypothesis is consistent with the eastward direction of prevalent horizontal winds predicted by MCD at all altitudes below 70 km within Atlantis Chaos. Since significant dust surface reservoirs are associated to low values of thermal inertia (Christiansen,1986), in Figure 9 we show the location of the storm in the context of a map of the thermal inertia in Atlantis Chaos, obtained from the data of Christensen and Moore (1992). The map shows that a large region of low thermal inertia is found at the west and north-west ridge of the Atlantis Chaos, which is therefore the closest candidate region as dust source.

On the other hand, MCD indicates a faint downward vertical wind component, slightly larger than the falling sedimentation velocity for the retrieved particle sizes and, hence, does not provide any clue on the possible dust lifting mechanism. We speculate that, after a local near-surface process triggers the dust lifting (possibly increasing the upward air motion locally through enhanced convection), eastward and downward MCD winds contribute to the dust accumulation and confinement towards the eastern ridge of Atlantis Chaos (see Figure 8).

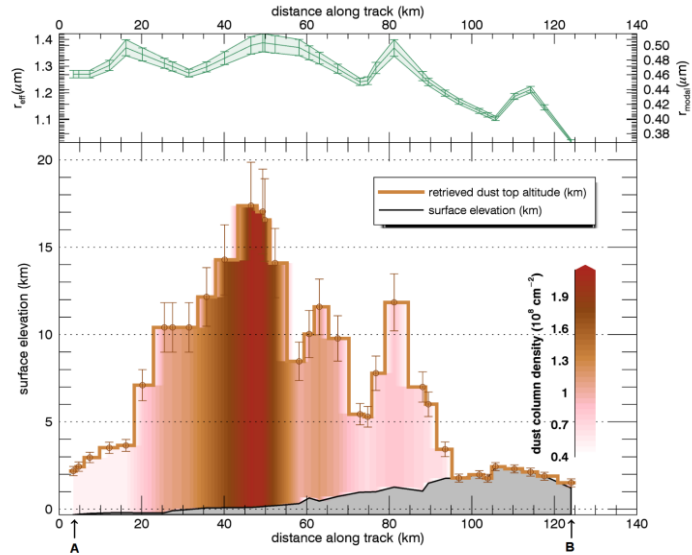

**Figure 8: transverse section of the structure of the dust storm along the track indicated by the black dashed line in panel *a* of Figure 7. A and B below the abscissa indicate the two extremes of the profile. The storm top altitude in km is shown as a thick brown line, filled in colours related to the dust column density, as indicated by the embedded colour bar. The corresponding altimetry profile at the base of the storm is shown filled in grey. The top panel shows the corresponding profile of particles dimensions as effective radii (left axis) and modal radii (right axis).**

Such framework is compatible with the results shown in Figure 7. In particular, now we focus on panel *d*, showing the dust column density of the storm. In Section 4.1 we have already seen how the northern arm of the cloud, straddling the crater's north-east ridge, is characterized by a decreased column density with respect to the southern one. This may happen due to either a reduced dust content or to a reduced dust column height (namely the difference between the cloud top altitude and the altimetry). However, the dust column height is not changing between the two arms and, hence, the reduced dust column content on the northern arm indicates less particles in that area. An increasing eastward-downward wind speed from the cloud centre to the ridge (MCD) could be the explanation of why we see less particles there.

At this point we guess where the dust is going to land once the storm has ended. To understand this, we make some considerations on the detectability of the surface mafic spectral signatures.

From the spectra shown in Figure 1, it is evident how the dust of the cloud is capable of masking these signatures when the optical depth is sufficiently high. We verified that this happens with a deposited dust thickness of more than 1 μm of dust on the surface (Figure 7f) corresponding to $\tau_{9.3} \geq$ 1. Therefore, if we presume that the cloud collapsed vertically over the region it occupies in orbit 1441_5, we would not be able to see the mafic surface spectral signatures when the storm ends. However, we verified that the spectra of the same region relative to orbits subsequent to the local storm (see Table 1) show such signatures.

For these reasons, we conclude that, after the storm has ended, the dust of the cloud has probably spread and deposited over a region larger than the brown line in Figure 9. Indeed, the region covered by the storm in orbit 1441_5 spans an area of about 30000 km$^2$ (see Section 3). On the other hand, the area of Atlantis Chaos is more than 10$^6$ km$^2$ wide. Hence, if the total retrieved dust deposited on the whole Atlantis Chaos area, its precipitable thickness would be of about 0.1 μm, ensuring the surface spectral signatures detectability.

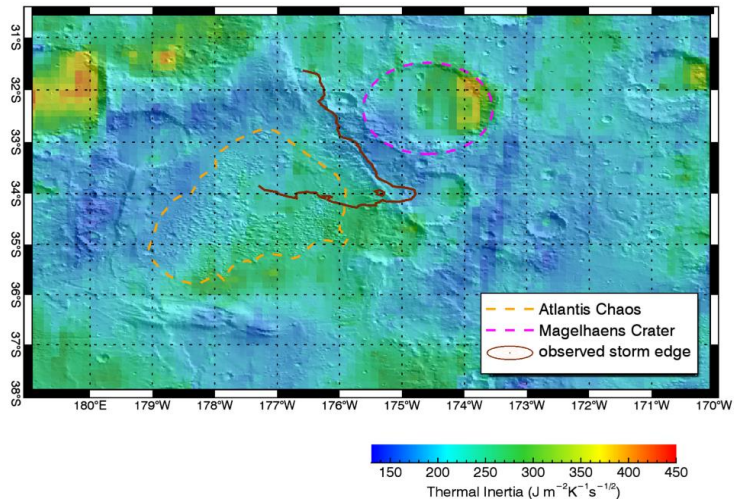

**Figure 9: thermal inertia map of Atlantis Chaos and neighbouring regions (data mosaic provided by JMARS software, Christensen et al., 2009).**

We now compare the outcome of our retrieval with the results relative to another local storm event. Our results for *ta* and $\tau_{9.3}$ are consistent with those of Määttänen et al. (2009), who studied a local storm observed by OMEGA at latitude 3° S and longitude 24.7° E (Ls = 135°, MY27), outside of a small crater northwards of the bigger Pollack crater. For that particular event, the authors provide a top altitude of the dust layer between 10 and 20 km, while we find a maximum of 18 km at the storm center. Moreover, they find a maximum optical depth of 10 at 1 μm. Our maximum $\tau_{9.3}$ is about 7 ± 2 at the center of the storm (see Table 2). The corresponding optical depth at 1 μm *($\tau_{1.0}$)* is about 14 ± 4 and, hence, is in agreement with the estimate from Määttänen et al. (2009). This comparison shows that, while the topography may have a confinement role as we suggested above, it is not affecting the microphysical properties and the vertical structure of these local dust events. Indeed, the same properties are found for our local storm, observed in the Atlantis Chaos crater, and for the Määttänen et al. (2009) storm, observed outside of another crater.

## 6. Summary and conclusions

In this study we retrieved maps of effective radius ($r_{eff}$), optical depth at 9.3 μm ($\tau_{9.3}$), and top altitude (*ta*) of a local dust storm in Atlantis Chaos, imaged by OMEGA in 2005 (see Section 2). Our results (Section 4) show that the cloud is characterized by an optically thick central region ($\tau_{9.3}$ > 7) where large particles ($r_{eff}$ = 1.6) are gathered at high altitude (*ta* > 18 km). Away from the centre, the cloud splits in two arms. The top altitude of the northern one exceeds 12 km and is characterized by a reduced density with respect to the southern one, where the top altitude is less than 10 km (see Figure 7). Moreover, it is evident from our results and from the data, that the topography plays a role in confining the cloud in the Atlantis Chaos region. Indeed, all the retrieved parameters undergo a sudden drop outside the borders of the crater, indicating that the cloud is not extending beyond them.

We debated the origin of the dust making the cloud (see Section 5). In dust free conditions, OMEGA observations show that the surface is characterized by mafic composition in good agreement with the thermal inertia values of the region (Mellon et al., 2000). Since low thermal inertia regions are renowned to be dust deposits (Christensen, 1986), we speculate that a region with low values of thermal inertia observed at the western border of Atlantis Chaos is the source of the storm. The combination of MCD horizontal and vertical winds is in agreement this hypothesis, suggesting that the dust, once lifted at west, is pushed eastward and downward by the winds, resulting in the confinement we observe at the north-east ridge of Atlantis Chaos. On the other hand, MCD vertical wind profiles alone do not provide any suggestion on the possible lifting mechanisms of dust.

We verified that when the storm ended the dust has probably deposited on an area larger than the one it occupies in orbit 1441_5. Indeed, if the cloud collapsed vertically on the ground, the thickness of the precipitable dust layer would prevent the detectability of the mafic absorption features of the surface. However, these features are present in observations of Atlantis Chaos registered by OMEGA after the dust storm.

**Acknowledgements**

This study has been performed within the UPWARDS project and funded in the context of the European Union's Horizon 2020 Programme (H2020-Compet-08-2014), grant agreement UPWARDS-633127. Moreover, the study has been co-funded within the PRIN INAF 2014 project. We would like to thank the whole OMEGA team for the support in this work.